\documentclass[nofootinbib,a4paper,aps,pra,showpacs,preprintnumbers,twocolumn]{revtex4-1}
\usepackage{cmap}
\usepackage{graphics,graphicx,epsfig}
\usepackage{epstopdf}
\usepackage[centertags]{amsmath}
\usepackage{amsfonts}
\usepackage{amssymb}
\usepackage{amsthm,color}
\usepackage{euscript}
\DeclareGraphicsExtensions{.pdf,.png,.jpg,.eps}
\usepackage[T1]{fontenc}
\usepackage[utf8]{inputenc}
\usepackage{xcolor}
\usepackage{hyperref}

\definecolor{linkcolor}{HTML}{000000} 
\definecolor{urlcolor}{HTML}{000000} 
\hypersetup{pdfstartview=FitH, linkcolor=linkcolor,urlcolor=urlcolor, colorlinks=true}

\usepackage[english]{babel}

\begin{document}
  
\title{Error of an arbitrary single-mode Gaussian transformation on a weighted cluster state using a cubic phase gate}
\author{E.R. Zinatullin$^1$}
\author{S.B. Korolev$^1$}
\author{A.D. Manukhova$^2$}
\author{T.Yu. Golubeva$^1$}
\affiliation{$^1$St. Petersburg State University, Universitetskaya nab. 7/9, St. Petersburg, 199034, Russia \\
$^2$Department of Optics, Palack\'y University, 17. Listopadu 12, Olomouc, 771 46, Czech Republic}
\begin{abstract}
In this paper, 
we propose two strategies for decreasing the error of arbitrary single-mode Gaussian transformations implemented using one-way quantum computation on a four-node linear cluster state. 
We show that it is possible to minimize the error of the arbitrary single-mode Gaussian transformation by a proper choice of the weight coefficients of the cluster state. 
We modify the computation scheme by adding a non-Gaussian state obtained using a cubic phase gate as one of the nodes of the cluster. 
This further decreases the computation error. 
We evaluate the efficiencies of the proposed optimization schemes comparing the probabilities of the error correction of the quantum computations with and without optimizations.
We have shown that for some transformations, the error probability can be reduced by up to 900 times.
\end{abstract}
\maketitle


\section{Introduction}

The main goal of quantum information and quantum computation is so-called universal quantum computation. 
This universal quantum computation can effect any unitary transformation over a finite number of variables to any degree of precision \cite{Lloyd} using the repeated application of local operations (affecting only a few variables at the same time).
One-way quantum computation \cite{Menicucci,Raussendorf,Nielsen} is a promising model for universal quantum computation. 
In our work, we will discuss the continuous variable one-way quantum computation technique \cite{Menicucci}. 
Unlike the discrete quantum systems, the use of continuous variables allows one to build schemes that give a significant measurement result each time they are addressed (deterministic circuits).
Moreover, such systems have great potential in terms of their scalability \cite{Yokoyama2013,Roslund,Chen,Yoshikawa,Larsen2019,Asavanant2019}. 
To achieve the universality of quantum computation, with continuous variables, it is necessary to be able to implement an arbitrary single-mode Gaussian (linear) transformation, a two-mode transformation, and at least one non-Gaussian (nonlinear) transformation \cite{Lloyd}.

The main resources of one-way quantum computation are cluster states. These, states belong to the family of highly entangled multipartite quantum states. Such states can be effectively parameterized by a mathematical graph \cite{Hein}. There are many ways to implement cluster states in continuous variables. It can be implemented with optomechanical systems \cite{Houhou}, atomic ensembles \cite{Sun}, hybrid variables \cite{Gabriel}, mixed (atomic-field) systems \cite{Tikhonov,Milne} and light fields \cite{Menicucci2,Yukawa1,Zhang,Ferrini,Medeiros}.

In continuous variables, cluster states are generated via a set of squeezed oscillators. In the idealized case, when the fluctuations in the squeezed quadrature are completely suppressed, the operations in the considered model are performed without errors. However, in reality, it is impossible to obtain an ideal squeezed state, oscillators with finite squeezing are used to generate a cluster. As a result, noises from non-ideal squeezed quadratures distort the result of operations and lead to the appearance of inherent errors. The presence of these errors is the main limiting factor for the model under consideration.
At the moment, the experimentally feasible squeezing is insufficient for performing fault-tolerant universal one-way quantum computations.
The achieved maximum is $-15$ dB \cite{Vahlbruch}, whereas the minimum required value for such computations (without using surface codes and the post-selection procedure) is  $-20.5$ dB \cite{Menicucci1}.

There are various methods to get around the limitations associated with insufficient squeezing. Such methods include the use of postselection \cite{Fukui} and surface codes \cite{Fukui,Noh,Fukui1,Noh1,Larsen1,Bourassa,Tzitrin}. For example, in \cite{Fukui}, the authors proposed a computation scheme that makes it possible to reduce the squeezing requirements to $-10.8$ dB. However, the main efforts are usually directed at error correction, whilst we propose to modify the computation scheme itself.
The resource state requirement can be lowered by using computation schemes less sensitive to the inherent error. 
The idea of constructing such schemes is to analyze the computational procedure to identify the nodes giving the noisiest result and reduce their influence.
Reducing errors in just one of the classes operations necessary for universal quantum computation helps to reduce the requirements for the resource state for the entire scheme.

The main goal of our work is to reduce arbitrary single-mode Gaussian operation errors. 
The first strategy is to employ the Gaussian transformations and to vary the weight coefficients of the cluster state used as a resource for the quantum computation. 
In \cite{Zinatullin1}, based on the quantum teleportation protocol, we showed that it is possible to decrease the quantum signal transmission error by using the weighted Controlled-Z (CZ) transformation  \cite{Larsen,Su2018,Alexander} as an entanglement operation. 
Increasing the weight coefficients made it possible to significantly decrease the error in one of the quadratures,
whilst the transformation maintains Gaussian.
Therefore, in Sec. \ref{Sgaus}, 
we study the impact of the cluster state weight coefficients on the single-mode operation errors.

It is necessary to mention that according to the No-Go theorem \cite{NoGo} Gaussian states cannot be used to correct Gaussian errors (determined by Gaussian transformations) in Gaussian states. 
The proposed method does not contradict this theorem. 
We reduce the computation error not by additional error correction, but by reducing the impact of the nodes that contribute the most errors.

The second strategy for decreasing the error is to use the clusters with non-Gaussian nodes. 
In \cite{Zinatullin2}, we have shown that it is possible to reduce the errors in the teleportation protocol by using the states prepared with the cubic phase gate \cite{GKP}. 
The teleportation protocol underlies the one-way quantum computation. 
Therefore, we apply the strategy of cluster modification to perform computation with fewer errors.
At the same time, we place the emphasis on the minimum change in the resource state (modifying only one node) which would lead to a decrease of the maximum error.
It is worth noting that the chosen non-Gaussian operation can be performed deterministically, when each measurement leads to the desired result. 
This is important for the scalability of quantum computation schemes and is advantageous for computations with continuous variables. 
Probabilistic procedures (such as the photon subtraction \cite{Opatrny,Cochrane}) would deprive the protocol of this advantage.

In \cite{Zinatullin2}, we proposed a strategy for decreasing the quantum teleportation protocol error by using a cubic phase gate to prepare a non-Gaussian resource state. 
In the Sec. \ref{Snongaus}, we apply this strategy to decrease the error of arbitrary single-mode Gaussian transformations. 
Noting that the generation of cubic phase states is also a challenging experimental problem. 
The first idea of cubic phase state generation was proposed by Gottesman, Kitaev, and Preskill in 2001 \cite{GKP,Ghose,Gu}. 
It turned out that this idea is difficult to implement. It requires performing the quadrature displacement operation by a value far from what is achievable in an experiment.
The cubic phase gate has long been an abstract mathematical transformation not realisable.
This situation has changed in recent years. 
Many works have been devoted to methods for cubic phase state generation \cite{Yukawa,YZhang,Asavanant} and the implementation of the cubic phase gate \cite{Hillmann,Marshall,Miyata,Yanagimoto,Konno} phenomena.
Particularly significant progress was achieved in the microwave frequency range - it was in this range that the cubic phase state was generated for the first time \cite{Kudra}. 
As a result, the cubic phase gate gradually turns from a purely theoretical transformation into a real-life device.

The paper is organized as follows. 
In Sec. \ref{Sgaus}, we describe the transformation scheme on a weighted four-node linear cluster state and demonstrate its arbitrariness for any values of weight coefficients. 
Also in this section, we estimate the errors in the considered scheme and proposed options for optimizing the transformation errors for experimentally achievable values of the weight coefficients. 
In Sec. \ref{Snongaus}, we study a modified computation scheme in which a cubic phase gate is used to prepare the cluster state, estimate the errors in this scheme, and perform their optimization. 
In Sec. \ref{Seff}, we evaluate the efficiency of optimization of the computation proposed in the previous sections based on the calculation of the error correction probability.
  

\section{Arbitrary single-mode Gaussian operation on a four-node cluster} \label{Sgaus}

\subsection{Transformation scheme on a linear four-node weighted cluster}

The principle of performing an arbitrary single-mode Gaussian operations on an unweighted linear four-node cluster is well known \cite{Gu,Ukai}. In this subsection, we will repeat similar transformations on a weighted cluster, and in the next subsection, we will demonstrate their arbitrariness for any values of the weight coefficients.

\begin{figure*}[t]
\begin{center}
\includegraphics[width=175mm]{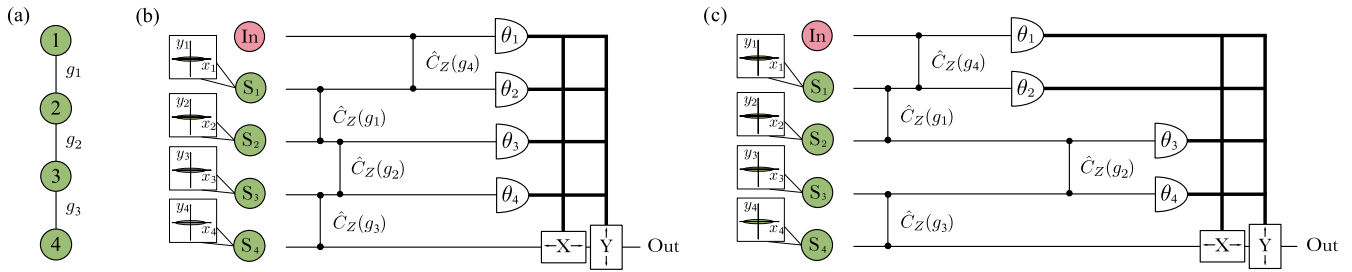}
\caption{(a) The configuration of the cluster state used as a resource for computation. (b) Scheme for implementing of the arbitrary single-mode Gaussian operation on a linear weighted four-node cluster state. (c) Scheme of implementation of arbitrary single-mode Gaussian operation on a pair of two-node cluster states. In the diagram: In is the input state; S$_j$ are squeezed states; $\hat C_z(g_j)$ is the CZ transformation with weight coefficient $g_j$; $\theta_j$ are the phases of the local oscillators employed for a balanced homodyne detection; and X and Y are operations that displace the corresponding quadratures of the fields in the channel, depending on the detection results.}
\label{shem1}
\end{center}
\end{figure*}

To get an explicit expression for the transformation error, let us start with constructing the cluster state.
The linear cluster state (Fig. \ref{shem1}(a)) is prepared from four oscillators squeezed in the $y$-quadrature. 
The quadratures of the $j$-th oscillator are described as 
\begin{align}
\hat x_{s,j}&=e^r \hat x_{0,j}, \qquad \hat y_{s,j}=e^{-r} \hat y_{0,j},
\label{resurs}
\end{align}
where $r$ is the squeezing coefficient, and $\hat x_{0,j}$ and $\hat y_{0,j}$ are the quadratures of the $j$-th oscillator in the vacuum state. 
The entanglement of cluster nodes with each other, as well as the entanglement of an input state with the node of the cluster state, will be carried out by using the CZ gate with the weight coefficients $g_{jk}$.
This transformation acts on the oscillators $j$ and $k$ as
\begin{align}
\hat C_{z}(g_{jk})=e^{2i g_{jk} \hat x_j \hat x_k}.
\end{align}
The weight coefficient $g_{jk}$ of the CZ gate can take any real value. The value of the weight coefficient determines the strength of the entanglement, i.e., how much information about the $j$-th system after the entanglement procedure will be contained in the $k$-th and vice versa. The sign of the weight coefficient indicates the creation of positive or negative correlations (anti-correlations) between oscillators. Hereinafter, in the paper, we consider positive weight coefficients, bearing in mind that their sign does not influence the error decrease.

All CZ transformations commute with each other. Thus, we can consider the computation on a four-node cluster state (see Fig. \ref{shem1}(b)) as a computation on a pair of two-node cluster states (see Fig. \ref{shem1}(c)). It simplifies the analysis of the scheme.

Let us consider the transformation performed on the first pair of resource states. The first and second squeezed oscillators are entangled using the CZ gate with the weight coefficient $g_1$. 
The input state is entangled to the first resource oscillator by a similar operation with a weight coefficient $g_4$. As the result, the amplitudes of the oscillators take the following form:
\begin{align}
&\hat a_{in}'=\hat x_{in}+i (\hat y_{in}+g_4 \hat x_{s,1}),\\
&\hat a_1'=\hat x_{s,1}+i\big(\hat y_{s,1}+g_1\hat x_{s,2}+g_4 \hat x_{in}\big),\\
&\hat a_2'=\hat x_{s,2}+i(\hat y_{s,2}+g_1 \hat x_{s,1}).
\end{align}
We then perform homodyne measurements with the local oscillator’s phases $\theta_1$ and $\theta_2$ over the input and first oscillators, respectively. It leads to the following equalities for the photocurrent operators:
\begin{align}
&\hat i_{in}=\beta \sin\theta_1 (\hat y_{in}+g_4 \hat x_{s,1})+\beta \cos\theta_1 \hat x_{in},\\
&\hat i_1=\beta \sin\theta_2 \big(\hat y_{s,1}+g_1\hat x_{s,2}+g_4 \hat x_{in}\big)+\beta \cos\theta_2 \hat x_{s,1},
\end{align}
where $\beta$ is the amplitude of the homodyne detector’s local oscillator. Such measurements, due to the entanglement of the resource state, lead to a change in the quadrature components of the second oscillator:
\begin{align}
\hat x_2'=&\left(\frac{\cot\theta_1 \cot\theta_2}{g_1 g_4}-\frac{g_4}{g_1}\right)\hat x_{in}+\frac{\cot\theta_2}{g_1 g_4}\hat y_{in}
-\frac{\hat y_{s,1}}{g_1} \nonumber\\
&+\frac{i_{1,m}}{\beta g_1 \sin\theta_2}-\frac{i_{in,m} \cot\theta_2}{\beta g_1 g_4 \sin\theta_1}, \label{x'2_1}\\
\hat y_2'=&-\frac{g_1 \cot\theta_1}{g_4}\hat x_{in}-\frac{g_1}{g_4}\hat y_{in}+\hat y_{s,2}+\frac{i_{in,m} g_1}{\beta g_4 \sin\theta_1}. \label{y'2_1}
\end{align}
Here, we replaced the operators of photocurrents with c-numbers corresponding to the results of the given measurement: $i_{1,m}$ and $i_{in,m}$. Let us rewrite this transformation in a matrix form:
\begin{align}
\begin{pmatrix}
\hat x'_2\\
\hat y'_2
\end{pmatrix}=&
\begin{pmatrix}
\frac{\cot\theta_1 \cot\theta_2}{g_1 g_4}-\frac{g_4}{g_1} & \frac{\cot\theta_2}{g_1 g_4}\\
-\frac{g_1 \cot\theta_1}{g_4} & -\frac{g_1}{g_4}
\end{pmatrix}
\begin{pmatrix}
\hat{x}_{in}\\
\hat{y}_{in}
\end{pmatrix}+
\begin{pmatrix}
-\frac{\hat{y}_{s,1}}{g_1}\\
\hat{y}_{s,2}
\end{pmatrix} \nonumber\\
&+
\begin{pmatrix}
\frac{i_{1,m}}{\beta g_1 \sin\theta_2}-\frac{i_{in,m} \cot\theta_2}{\beta g_1 g_2 \sin\theta_1}\\
\frac{i_{in,m} g_1}{\beta g_2 \sin\theta_1}
\end{pmatrix}.
\label{x2}
\end{align}
It is well known that this Gaussian transformation is not arbitrary for the unit weight coefficients. 
To ensure arbitrariness, it is necessary to perform a similar operation again on another pair of nodes.
Since our goal is to ensure arbitrariness for any $g_j$, 
and for $g_j=1$ it is not arbitrary,
the scheme needs to be complemented.

The operation on the second pair of nodes, up to weight coefficients of the CZ gate, repeats the operation on the first pair of nodes. Thus, the second part of the scheme operates similar to the transformation (\ref{x2}), where the input data are the quadratures $x_2'$ and $y_2'$. At the output of the scheme, the c-number components of the quadratures of the field are compensated by displacement, depending on the values of the measured photocurrents. We introduce a new notation:
\begin{align}
\cot\theta_2'=\frac{\cot\theta_2}{g_4^2}, \qquad \cot\theta_4'=\frac{\cot\theta_4}{g_2^2}.
\label{theta_new}
\end{align}
Note that in the new notation, the mathematical expression for the input-output transformation will depend not on the weight coefficients themselves, but on their ratio. It is convenient for further analysis. Thus, the operation carried out by our scheme has the form 
\begin{align}
\begin{pmatrix}
\hat x_{out}\\
\hat y_{out}
\end{pmatrix}=&
U(\theta_1,\theta_2',\theta_3,\theta_4')
\begin{pmatrix}
\hat{x}_{in}\\
\hat{y}_{in}
\end{pmatrix}
+
\delta \hat {\bf e}_0(\theta_3,\theta_4').
\label{out0}
\end{align}
Here, the desired transformation performed on the input state is described by the matrix
\begin{align}
U(\theta_1,\theta_2',\theta_3,\theta_4')=&
\begin{pmatrix}
\frac{\cot\theta_3 \cot\theta_4'-1}{g_3/g_2} & \frac{\cot\theta_4'}{g_3/g_2}\\
-\frac{g_3 \cot\theta_3}{g_2} & -\frac{g_3}{g_2}
\end{pmatrix} \nonumber\\
&\times
\begin{pmatrix}
\frac{\cot\theta_1 \cot\theta_2'-1}{g_1/g_4} & \frac{\cot\theta_2'}{g_1/g_4}\\
-\frac{g_1 \cot\theta_1}{g_4} & -\frac{g_1}{g_4}
\end{pmatrix},
\label{matrix_U}
\end{align}
and the transformation error associated with the finite squeezing of the oscillators is described by the vector
\begin{align}
\delta \hat {\bf e}_0(\theta_3,\theta_4')=&
\begin{pmatrix}
\frac{\cot\theta_3 \cot\theta_4'-1}{g_3/g_2} & \frac{\cot\theta_4'}{g_3/g_2}\\
-\frac{g_3 \cot\theta_3}{g_2} & -\frac{g_3}{g_2}
\end{pmatrix}
\begin{pmatrix}
-\frac{\hat{y}_{s,1}}{g_1}\\
\hat{y}_{s,2}
\end{pmatrix} +
\begin{pmatrix}
-\frac{\hat{y}_{s,3}}{g_3}\\
\hat{y}_{s,4}
\end{pmatrix}.
\label{de_0}
\end{align}
It should be noted that in the scheme under consideration, the transformation error $\delta \hat {\bf e}_0$ depends only on the angles $\theta_3$ and $\theta_4'$. This is because measuring the second pair of resource oscillators transforms the error from the first pair of resource oscillators.

\subsection{Arbitrariness of the transformation with arbitrary weight coefficients}

First, we need to find out if the single-mode Gaussian transformation $U$ is arbitrary. It was shown in \cite{Ukai} that the $U$ will be arbitrary when weight coefficients of the CZ gate are unity. However, we need to check whether arbitrariness is preserved for arbitrary non-unity weight coefficients. To do this, we will show that it is possible to choose the phases of local oscillators of homodyne detectors in such a way that the matrix $U(\theta_1,\theta_2',\theta_3,\theta_4')$ is equal to any given arbitrary symplectic matrix, i.e.
\begin{align}
U(\theta_1,\theta_2',\theta_3,\theta_4')=
\begin{pmatrix}
a & b\\
c & d
\end{pmatrix},
\label{univer1}
\end{align}
where the coefficients of matrix satisfy the condition
\begin{align}
ad-bc=1. \label{univer_usl}
\end{align}

Taking into account the explicit form (\ref{matrix_U}) of the matrix $U$, the Eq. (\ref{univer1}) is equivalent to the system of equations 
\begin{align}
&\frac{g_2 g_4}{g_1 g_3} (\cot\theta_1 \cot\theta_2'-1)(\cot\theta_3 \cot\theta_4'-1) \nonumber\\
&\qquad -\frac{g_1 g_2}{g_3 g_4} \cot\theta_1\cot\theta_4'=a, \label{eq_a}\\
&\frac{g_2 g_4}{g_1 g_3} \cot\theta_2'(\cot\theta_3 \cot\theta_4'-1)-\frac{g_1 g_2}{g_3 g_4} \cot\theta_4'=b,\\
&-\frac{g_3 g_4}{g_1 g_2} (\cot\theta_1 \cot\theta_2'-1) \cot\theta_3+\frac{g_1 g_3}{g_2 g_4} \cot\theta_1=c,\\
&-\frac{g_3 g_4}{g_1 g_2} \cot\theta_2' \cot\theta_3+\frac{g_1 g_3}{g_2 g_4}=d. \label{eq_d}
\end{align}
Due to the condition (\ref{univer_usl}) on the matrix coefficients, any one of these equations can be derived from the three remaining. Thus, there are only three independent equations in four variables. The system of Eqs. (\ref{eq_a})--(\ref{eq_d}) is not uniquely solved, and one of the phases can be chosen as a free parameter, which we can change. We can see that for $\theta_2'=\pi/2$ or $\theta_3=\pi/2$ the Eq. (\ref{eq_d}) turns into equality so that we lose the ability to solve the system for an arbitrary matrix. Therefore, the phases $\theta_2'$ and $\theta_3$ cannot be chosen as a free parameter. Of the remaining two phases, the phase $\theta'_4$ is the best candidate to be a free parameter, since the transformation error depends on it. In the future, with the right choice of the phase $\theta_4'$, we will be able to minimize the transformation error.

For some arbitrary fixed value of phase $\theta_4'$, the solution of the system (\ref{eq_a})--(\ref{eq_d}) exists for the remaining phases and has the form
\begin{align}
&\cot\theta_1=\frac{c}{d}+\frac{\left(\frac{g_1 g_3}{g_2 g_4}-d\right)\frac{g_3}{g_2}}{\left(\frac{g_3^2}{g_2^2}b+d \cot\theta_4'\right)\frac{g_1}{g_4}d}, \label{eq_1}\\
&\cot\theta_2'=-\frac{g_1 g_2}{g_3 g_4}\left(\frac{g_3^2}{g_2^2}b+d \cot\theta_4'\right),\\
&\cot\theta_3=\frac{d-\frac{g_1 g_3}{g_2 g_4}}{\frac{g_3^2}{g_2^2}b+d \cot\theta_4'}. \label{eq_3}
\end{align}
Thus, for any arbitrary fixed value $\theta_4'$, we can choose the phases of the local oscillators $\theta_1$, $\theta_2'$, and $\theta_3$ in such a way that the matrix (\ref{matrix_U}) is equal to any given arbitrary symplectic matrix. This means that the single-mode Gaussian transformation given by the matrix (\ref{matrix_U}) is arbitrary.

\subsection{Single mode transformation error on a weighted cluster}

\subsubsection{Optimization for arbitrary values of weight coefficients}

Let us estimate the errors in the considered scheme of one-way computations. To do this, we pass from the error vector to one consisting of variances $\langle \delta \hat {\bf e}^2_0 \rangle$. We assume resource oscillators to be statistically independent and squeezed equally, i.e. $\langle \hat y^2_{s,j} \rangle \equiv \langle \delta \hat y^2_s \rangle$ for $j\in1,2,3,4$. Then, the vector of variances has the form:
\begin{align}
\langle \delta \hat {\bf e}^2_0 \rangle=
\begin{pmatrix}
\frac{1}{g_1^2}\left(\frac{\cot\theta_3 \cot\theta_4'-1}{g_3/g_2}\right)^2+\left(\frac{\cot\theta_4'}{g_3/g_2}\right)^2+\frac{1}{g_3^2}\\
\frac{1}{g_1^2}\left(\frac{g_3 \cot\theta_3}{g_2}\right)^2+\frac{g_3^2}{g_2^2}+1
\end{pmatrix}
\langle \delta \hat y^2_s \rangle.
\label{de_0_2}
\end{align}
Note that the transformation errors directly depend on the weight coefficients $g_1$ and $g_3$. On the other hand, the ratio between weight coefficients $g_3/g_2$ determines which type of operation at certain phase values is performed. Therefore, it makes sense to compare the transformation errors only for a fixed ratio $g_3/g_2$, i.e. to compare the errors of the same operations. To achieve it, let us substitute solutions (\ref{eq_1})--(\ref{eq_3}) into Eq. (\ref{de_0_2}). As a result, we get the error variance vector:
\begin{align}
\langle \delta \hat {\bf e}^2_0 \rangle=&
\begin{pmatrix}
\frac{1}{g_3^2}+\left(\frac{g_2}{g_3}\right)^2 \cot ^2 \theta_4'+\frac{{g_2}^2 \left({b} \frac{g_4}{g_1}+\frac{g_2}{g_3}\cot \theta_4'\right)^2}{g_3^2g_4^2\left({b}+{d}\left(\frac{g_2}{g_3}\right)^2\cot \theta_4'\right)^2}\\
1+\frac{1}{\left(\frac{g_2}{g_3}\right)^2}+\frac{\left({d} \frac{g_2}{g_3} \frac{g_4}{g_1}-1\right)^2}{g_4^2 \left({b}+{d} \left(\frac{g_2}{g_3}\right)^2\cot \theta_4'\right)^2} 
\end{pmatrix}\nonumber\\
&\times\langle \delta \hat y^2_s \rangle. \label{error_eq}
\end{align}
This vector depends on the implemented operation (on the values of $d$ and $b$), on the phase $\theta_4'$ and on the value of the weight coefficients of the cluster state $g_1$, $g_2$, $g_3$, $g_4$.

The dependence of the error on the weight coefficients of the cluster state means that for each transformation (for each $b$ and $d$) there is a cluster state configuration that yields the minimal error. One can use this feature to construct non-universal quantum calculators capable of solving specific tasks. Here and below, we will be interested in the errors of universal quantum computation. Therefore, we are leaving out of consideration calculators for local problems (which are usually called quantum simulators). Unfortunately, in practice, when building universal computer, we cannot choose weight coefficients for each transformation, since this would require us to rebuild the cluster generation scheme each time. In reality, we have a cluster state with fixed weight coefficients. We need to choose the weight coefficients so that any transformation have a small error. Our goal is to identify such weight coefficients.

From the Eq. (\ref{error_eq}), one can see that if we impose the conditions on the weight coefficients: $g_1 \gg g_4$, $g_2 \gg g_3$, $g_3 \gg 1$, $g_4 \gg 1$, and set $\theta_4'$ equal to $\pi/2$, then the error will be proportional to the following vector:
\begin{align}
\langle \delta \hat {\bf e}^2_0 \rangle \approx
\begin{pmatrix}
0\\
1
\end{pmatrix}\langle \delta \hat y^2_s \rangle, \label{error_eq_1}
\end{align}
That is, we get the minimum computation error, which does not depend on the implemented operations (on $b$ and $d$). Unfortunately, in experiments, we cannot make the weight coefficients infinitely large, since it require infinite squeezed resource states. Let us see how large we can make them in practice.

\subsubsection{Experimental implementation of CZ transformation}

\begin{figure}[t]
    \centering
    \includegraphics[scale=1]{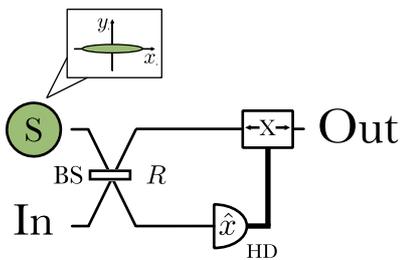}
    \caption{Implementation of in-line squeezing.}
    \label{fig:Squeezer}
\end{figure}
 
To understand what restrictions are imposed on the weight coefficients of the cluster state, let us consider the structure of the CZ  transformation. As is known, the CZ  gate with the weight coefficient $g$ transforms the vector of input quadratures into the vector of output quadratures according to the rule:
 \begin{align}
\begin{pmatrix}
\hat{X}_{out,1}\\
\hat{X}_{out,2}\\
\hat{Y}_{out,1}\\
\hat{Y}_{out,2}
\end{pmatrix}=
\begin{pmatrix}
1 & 0 & 0 & 0\\
0 & 1 & 0 & 0\\
0 & g & 1 & 0 \\
g & 0 & 0 & 1
\end{pmatrix}\begin{pmatrix}
\hat{x}_{in,1}\\
\hat{x}_{in,2}\\
\hat{y}_{in,1}\\
\hat{y}_{in,2}
\end{pmatrix}.
\end{align}
To understand how this transformation can be implemented experimentally, we need to use the Bloch-Messiah decomposition for the CZ matrix. It has the following form:
\begin{widetext}
\begin{align}\label{decomposition}
\begin{pmatrix}
1 & 0 & 0 & 0\\
0 & 1 & 0 & 0\\
0 & g & 1 & 0 \\
g & 0 & 0 & 1
\end{pmatrix}=\begin{pmatrix}
1 & 0 & 0 & 0\\
0 & 0 & 0 & -1\\
0 & 0  & 1 & 0 \\
0 & 1 & 0 & 0
\end{pmatrix}\begin{pmatrix}
t & r & 0 & 0\\
r & -t & 0 & 0\\
0 & 0  & t & r \\
0 & 0 & r & -t
\end{pmatrix}\begin{pmatrix}
\sqrt{s} & 0 & 0 & 0\\
0 & \frac{1}{\sqrt{s}} & 0 & 0\\
0 & 0  & \frac{1}{\sqrt{s}} & 0 \\
0 & 0 & 0 & \sqrt{s}
\end{pmatrix}\begin{pmatrix}
r & t & 0 & 0\\
t & -r & 0 & 0\\
0 & 0  & r & t \\
0 & 0 & t & -r
\end{pmatrix}\begin{pmatrix}
1 & 0 & 0 & 0\\
0 & 0 & 0 & 1\\
0 & 0  & 1 & 0 \\
0 & -1 & 0 & 0
\end{pmatrix},
\end{align}
\end{widetext} 
where
\begin{align} \label{coef}
    &r=\frac{\sqrt{s}}{\sqrt{1+s}}, \quad t=\frac{1}{\sqrt{1+s}}, \nonumber\\
    & s=\frac{1}{2}\left(2+g^2-g\sqrt{4+g^2}\right).
\end{align}
Here, the first and last matrices describe the phase shifters. The second and fourth matrices describe the beam splitter transformation, and the third matrix describes the squeezing. Since we consider the case of non-negative weight coefficients $g$, then $s \in \left[0,1\right]$.

The main difficulty in the practical implementation of the CZ gate is the in-line squeezing. The in-line squeezing is the squeezing transformation performed on the oscillator inside the computation scheme. For in-line squeezing of the oscillator in one of the quadratures, the scheme shown in Fig. \ref{fig:Squeezer} is usually applied. 
In this scheme, the squeezed state S is entangling with the input state In (the state we are transforming) on the beam splitter BS with a reflection coefficient $R$. Next, the $x$-quadrature of the state in the lower channel is measured using the HD homodyne detector. After that, the measurement result is sent to a device that displaces the quadratures of the state in the upper channel (the device is indicated X in the diagram) depending on the measurement result. The quadratures of the output state Out can be represented as:
\begin{align}
\begin{pmatrix}
\hat{X}_{out}\\
\hat{Y}_{out}
\end{pmatrix}=
\begin{pmatrix}
\frac{1}{\sqrt{R}} & 0 &\\
0 & \sqrt{R} 
\end{pmatrix}\begin{pmatrix}
\hat{x}_{in}\\
\hat{y}_{in}
\end{pmatrix}
+\begin{pmatrix}
0\\
\sqrt{1-R}\hat{y}_s
\end{pmatrix}.
\end{align}
It can be seen from this expression that the quadrature $\sqrt{R}\hat{y}_{in}$ is squeezed, since the reflection coefficient is in the range $R \in [0,1]$. Since the squeezing coefficient $s$ in the Eq. (\ref{coef}) is in the range $s \in [0,1]$,  we can set $s=R$, implying that the reflection coefficient is responsible for the squeezing. As we can see, the main bottleneck of this implementation of the squeezing transformation is that there is an error that is added to the computation results. This error is proportional to the squeezed quadrature of the auxiliary oscillator S. The more the quadrature is squeezed, the better the gate is realized.

We use the considered scheme as part of the CZ transformation. Let us estimate the squeezing of the auxiliary quantum oscillators required for this. For the error to remain small compared to the main transformation, the following requirement must be met:
\begin{align}
    (1-R)\langle \delta \hat{y}_s^2 \rangle \ll R\langle \delta \hat{y}_{in}^2\rangle
\end{align}
or
\begin{align} \label{eq_error_1}
    10\text{lg} \left[ 4\langle \delta \hat{y}_s^2 \rangle \right] \ll 10\text{lg}\left[\frac{4R}{1-R}\langle \delta \hat{y}_{in}^2\rangle \right].
\end{align}
In this section, for simplicity, we consider fluctuations of the coherent state ($\langle\delta \hat{y}_{in}^2\rangle=1/4$) as a variance of the input state.

\begin{figure}[t]
    \centering
    \includegraphics[scale=0.87]{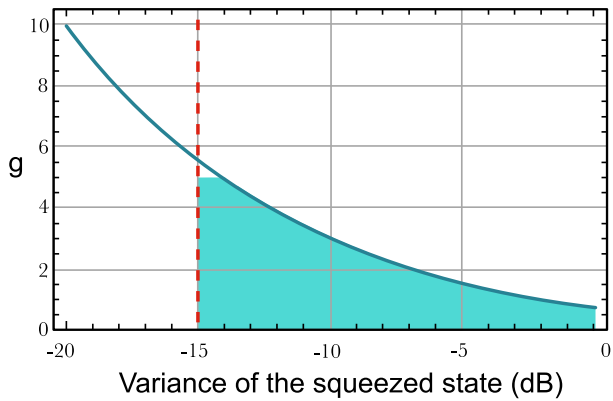}
    \caption{A graph of the dependence of the value of the CZ gate weight coefficient on the variance of the squeezed oscillator used to implement this transformation. On the graph: the blue solid line indicates the dependence of the weight coefficient on the variance of the squeezed state, the red dotted line indicates the squeezing limit experimentally implemented to date.}
\label{function_ru}
\end{figure}

To understand what kind of squeezing we can implement experimentally, we write the expression in a general form for an arbitrary $g$. To do this, we take into account equality $s=R$, and substitute the relationship (\ref{coef}) between the values of the squeezing ratio $s$ and the weight coefficient $g$ to the Eq. (\ref{eq_error_1}). As a result, the final dependence of the weight coefficient on the squeezing of the auxiliary oscillator can be estimated by the following inequality:
\begin{align} \label{squeez}
g < \frac{10^{-x/20}}{\sqrt{1 + 10^{x/10}}},
\end{align}
where $x=10\log_{10}\left(4\langle \delta \hat{y}^2_s\rangle \right)$. For greater clarity, the Eq. (\ref{squeez}) is shown in Fig. \ref{function_ru}. 
At the moment, the squeezing that has been experimentally demonstrated is $-15$ dB \cite{Vahlbruch}. Given the fact that the error should be small compared to the main transformation, we can say that the weight coefficient $g$ can be no more than $5$. This value we use further for numerical estimates, as corresponding to the maximum experimentally realized squeezing.

\subsubsection{Optimization of errors of one-way transformations for bounded values of $g$}

Let us compare the computation errors with different weight coefficients. We consider the $||\cdot||_{\infty}$ norm as a measure of errors. This norm has the form $||\langle \delta \hat {\bf e}^2_0 \rangle||_{\infty}=\max \left[\langle \delta \hat {\bf e}^2_0 \rangle _1,\langle \delta \hat {\bf e}^2_0 \rangle _2\right]$.

As we have discussed, to minimize the error, we need to require the conditions: $g_1 \gg g_4$, $g_2 \gg g_3$, $g_3 \gg 1$, $g_4 \gg 1$ and $\theta_4'=\pi/2$. From the first two inequalities, we can conclude that $g_1$ and $g_2$ should be chosen as maximum, i.e. $g_1=g_2=5$. It follows from the remaining conditions that $g_3$ and $g_4$ should be large enough (compared to unity), so we chose $g_3=g_4=4$. As before, we investigate the effect of weight coefficients on transformation errors for $\theta_4'=\pi/2$.

It should be noted that one can carry out a multidimensional optimization to find the global minimum of computation errors. However, this is a computationally difficult problem. In addition, the global minimum may lie outside the admissible weight coefficients. In this regard, we limited ourselves to the selection of weight coefficients of the cluster state, which provides a smaller error for a larger number of operations.

\begin{figure}[t]
\centering
\includegraphics[scale=0.7]{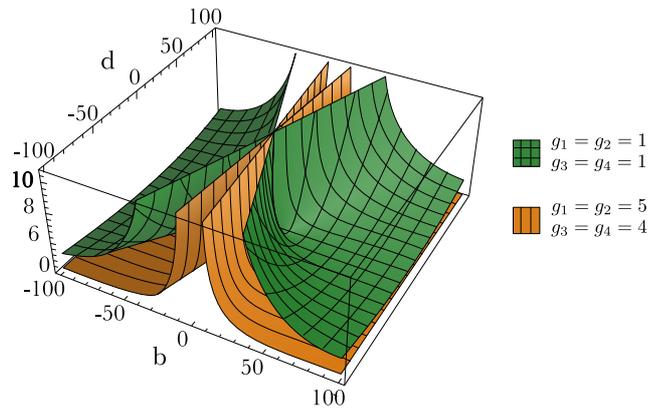}
\caption{The distribution of errors $||\langle \delta \hat {\bf e}^2_0 \rangle||_{\infty}/\langle \delta \hat y^2_s \rangle$ depending on the implemented single-mode transformation, i.e. it depending on b and d (see Eq. (\ref{univer1})). The graph shows two error surfaces corresponding to computations on two cluster states. The lower surface corresponds to the case of computations on a weighted optimized cluster state ($g_1=g_2=5$, $g_3=g_4=4$). The upper surface corresponds to the computational errors on the unweighted cluster state ($g_1=g_2=g_3=g_4=1$).}
\label{Fig_3D}
\end{figure}

\begin{figure*}[t]
\centering
\includegraphics[scale=0.50]{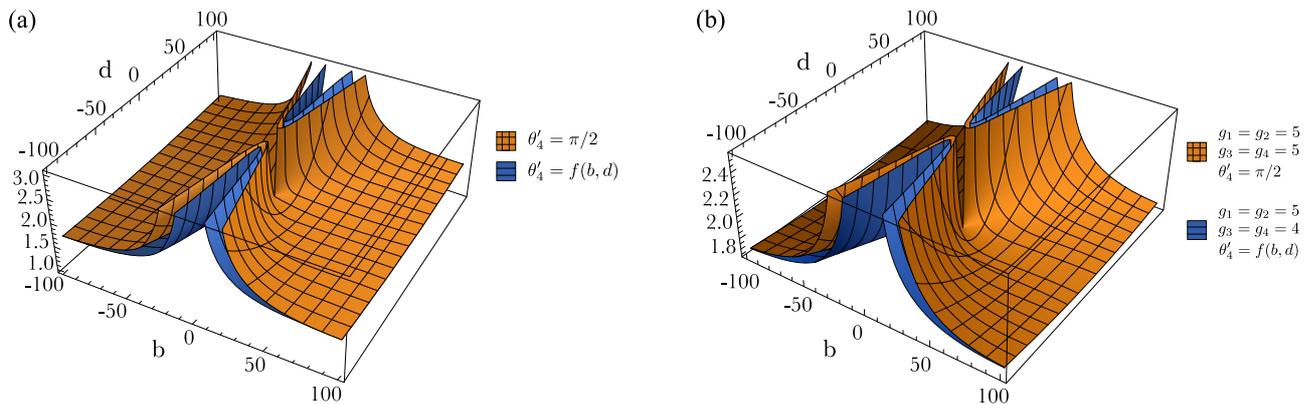}
\caption{The distribution of single-mode transformation errors $||\langle \delta \hat {\bf e}^2_0 \rangle||_{\infty}/\langle \delta \hat y^2_s \rangle$ depending on b and d (see Eq. (\ref{univer1})). 
(a) Transformations performed on the cluster state with optimized weight coefficients. On the diagram, the upper surface corresponds to the case when the phase of the homodyne detector is kept constant $\theta_4'=\pi/2$. The lower surface corresponds to the case when the phase is optimised for the performed transformation (for $b$ and $d$). 
(b) On the diagram, the upper surface corresponds to computations on the cluster state with weight coefficients $g_1=g_2=g_3=g_4=5$ and phase $\theta_4'=\pi/2$. Lower surface corresponds to computations on the cluster state with optimized weight coefficients $g_1=g_2=5$, $g_3=g_4=4$ and optimized homodyne detector phase $\theta_4'=f(b,d)$.}
\label{Fig_3D_2}
\end{figure*}

We first compare the errors for computation on a cluster state with the optimized weight coefficients ($g_1=g_2=5$ and $g_3=g_4=4$) with the case when an unweighted cluster state is used, i.e. a cluster state with unit weight coefficients ($g_1=g_2=g_3=g_4=1$). It is this state that is often considered by researchers as a universal state for implementing quantum Gaussian transformations \cite{Gu,Ukai}. Computation errors when using these two cluster states are shown in Fig. \ref{Fig_3D}. One can see that errors of computation on the weighted cluster state with optimized weight coefficients are always lower than on the unweighted cluster state. In other words, we have found that a weighted cluster state with optimized weight coefficients is better suited for the computation.

After we made the optimization of errors due to the weight coefficients, we can proceed to optimization due to the phase $\theta_4'$ of the homodyne detector. Until now, we have considered only the simplest case, when $\theta_4'=\pi/2$. As follows from Eq. (\ref{error_eq_1}), this case is optimal if the weight coefficients obey the conditions: $g_1 \gg g_4$, $g_2 \gg g_3$, $g_3 \gg 1$, $g_4 \gg 1$. As we found out, in reality, the weight coefficients are very limited in value. This limitation means that we cannot achieve the minimum error limit (\ref{error_eq_1}). This means that the value of phase $\theta_4'=\pi/2$ is not necessarily optimal. It follows from Eq. (\ref{error_eq}) that by selecting phase $\theta_4'$ for each specific transformation (for specific $b$ and $d$), we can minimize the errors. It is this optimization that was carried out by us below.
 
The optimization process consists in finding the minimum value of the function $||\langle \delta \hat {\bf e}^2_0 \rangle||_{\infty} =h(\theta_4',b,d)$ by the parameter $\theta_4'$ depending on $b$ and $d$.  As a result of the optimization, we obtain the dependence of the optimal phase on the operation, i.e.  $\theta_{4,min}'=f(b,d)$. Fig. \ref{Fig_3D_2}(a) shows the error surfaces of computations on the cluster state with weight coefficients $g_1=g_2=5$, $g_3=g_4=4$ when phase $\theta_4'=\pi/2$ and when phase is optimized depending on the operation ($\theta_4'=\theta_{4,min}'=f(b,d)$). It can be seen from the figure that the errors of single-mode transformation at optimized phase $\theta_4'$ are always smaller than the errors at $\theta_4'=\pi/2$. In other words, the error of any single mode operation can be further decreased by optimizing the phase of the homodyne detector.

To demonstrate the superiority of our optimized scheme (with optimized weight coefficients and optimized phase $\theta_4'$), let us compare it with the case of computations on a cluster state with the maximum weight coefficients ($g_1=g_2=g_3=g_4=5$) at $\theta_4'=\pi/2$. Fig. \ref{Fig_3D_2}(b) shows the errors $||\langle \delta \hat {\bf e}^2_0 \rangle||_{\infty}$ obtained by transformations in these two schemes. It can be seen from the graph that the error in the optimized case of computations is less than the error obtained when using the cluster with maximum weight coefficients. It is important to note that to create a cluster with large weight coefficients, we need to implement a squeezing transformation with a large coefficient $s$ (see Eq. (\ref{coef})). To perform such a transformation, we need an additional resource. Without loss of generality, we can say that additional energy is required. The larger the squeezing coefficient $s$, the more energy is needed. All this means that it takes more energy to create a cluster with maximum weight coefficients than to create an optimized cluster state. Thus, we can conclude that the smart use of the available physical resource (proper distribution of weight coefficients and smart choice of phases of the homodyne measurement) helps to reduce the quantum computation error.


\section{Single-mode operation using a cubic phase gate.} \label{Snongaus}

\subsection{Transformation scheme with a cubic phase gate}

As we have shown in the previous section, the single-mode Gaussian transformation on a linear 4-node weighted cluster state is arbitrary. Also, it is possible to reduce the error of this transformation by optimizing the weight coefficients of the cluster state. However, part of the operations still has significant errors.

\begin{figure*}[t]
\begin{center}
\includegraphics[width=155mm]{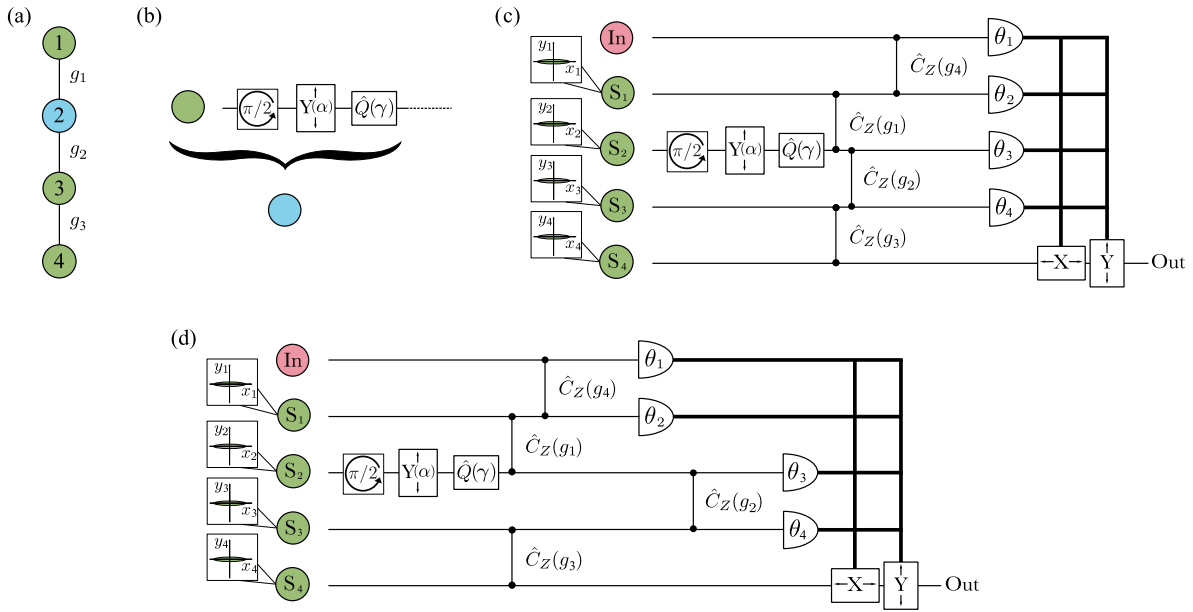}
\caption{(a) The configuration of the cluster state used as a resource for computation: Gaussian nodes are shown in green, non-Gaussian nodes are shown in blue. (b) The scheme of preparing a non-Gaussian resource state. (c) Scheme for implementing the arbitrary single-mode Gaussian operation on a linear weighted four-node cluster state using a cubic phase gate. (d) Scheme of implementation of arbitrary single-mode Gaussian operation on a pair of two-node cluster states using a cubic phase gate. In the diagram: Y$(\alpha)$ is operations that displace $y$-quadrature on a real value $\alpha$, $\hat Q(\gamma)$ is the cubic phase gate with nonlinearity $\gamma$.}
\label{shem2}
\end{center}
\end{figure*}

In article \cite{Zinatullin2} we have shown that it is possible to reduce the teleportation error by using the cubic phase gate to prepare a non-Gaussian resource state. In this section, we will apply this technique to decrease the error of the arbitrary single-mode Gaussian transformation scheme on a 4-node cluster state. To do it, we include the non-Gaussian state as the second node of the cluster (Fig. \ref{shem2}(a)). It can be seen from Eq. (\ref{de_0_2}) that the expression for the $x$-quadrature error has a term depending on the phase $\theta_4$. We cannot suppress this term by the CZ gate weight coefficients. It arises as a consequence of the finite squeezing of the second resource oscillator, which is why we replace the second node with a non-Gaussian resource.

As in the previous scheme, we use oscillators squeezed in $y$-quadrature as a resource for cluster state preparation. A non-Gaussian resource is prepared by sequentially applying the phase shift on $\pi/2$:
\begin{align}
\hat R_{2}(\pi/2)=e^{i \frac{\pi}{2} \hat a_2^\dagger \hat a_2},
\end{align}
$y$-quadrature displacement on $\alpha>0$:
\begin{align}
\hat Y_2(\alpha)=e^{2i \alpha \hat x_2},
\end{align}
and a cubic phase gate 
\begin{align}
\hat Q_2(\gamma)=e^{-2i\gamma \hat y_2^3}, \label{qfg}
\end{align}
where $\gamma$ is the coefficient of nonlinearity (see Fig. \ref{shem2}(b)). Thus, the second recourse oscillator will proceed to the non-Gaussian state, which is described by the following equation:
\begin{align}
\hat a_2=-\hat y_{s,2}+3\gamma(\alpha+\hat x_{s,2})^2+i(\alpha+\hat x_{s,2}).
\end{align}
The cubic phase gate (\ref{qfg}) deforms the uncertainty region of the squeezed in $x$-quadrature state in such a way that a parabola is formed on the phase plane. However, when we displace $y$-quadrature by a positive value $\alpha$ that satisfies condition $\alpha^2\gg\langle \hat x_{s,2}^2 \rangle$, the quadrature values of the second oscillator will lie in the first quadrant of the phase plane. In other words, only one of the branches of the parabola will remain on the phase plane.

As in the previous section, let us start with the analysis of the transformation performed on the first pair of resource states. The first Gaussian and second non-Gaussian resource oscillators are entangled using the CZ transformation with the weight coefficient $g_1$, and the input state is entangled with the first resource oscillator by the CZ gate with a weight coefficient $g_4$. After the entanglement, the amplitudes of the oscillators are described by the following equations:
\begin{align}
&\hat a_{in}'=\hat x_{in}+i (\hat y_{in}+g_4 \hat x_{s,1}),
\end{align}
\begin{align}
&\hat a_1'=\hat x_{s,1}+i\big(\hat y_{s,1}-g_1 \hat y_{s,2}+3 g_1 \gamma(\alpha+\hat x_{s,2})^2+g_4 \hat x_{in}\big),\\
&\hat a_2'=-\hat y_{s,2}+3\gamma(\alpha+\hat x_{s,2})^2+i(\alpha+\hat x_{s,2}+g_1 \hat x_{s,1}).
\end{align}
We can see that the first resource oscillator is now containing the nonlinearity from the non-Gaussian oscillator due to entanglement. We then perform homodyne measurements with the local oscillator’s phases $\theta_1$ and $\theta_2$ over the input and first resource oscillators, respectively. Thus, for the operators of photocurrent we get
\begin{align}
&\hat i_{in}=\beta \sin\theta_1 (\hat y_{in}+g_4 \hat x_{s,1})+\beta \cos\theta_1 \hat x_{in},\\
&\hat i_1=\beta \cos\theta_2 \hat x_{s,1}+\beta \sin\theta_2 \big(\hat y_{s,1}-g_1 \hat y_{s,2} \nonumber\\
&\qquad+3 g_1 \gamma(\alpha+\hat x_{s,2})^2+g_4 \hat x_{in}\big),
\end{align}
where $\beta$ is the amplitude of the homodyne detector’s local oscillator.
Due to the entanglement of the resource state, such a measurement  lead to a change in the quadrature components of the field in the second channel:
\begin{widetext}
\begin{align}
\hat x_2'=&\left(\frac{\cot\theta_1 \cot\theta_2}{g_1 g_4}-\frac{g_4}{g_1}\right)\hat x_{in}+\frac{\cot\theta_2}{g_1 g_4}\hat y_{in}
-\frac{\hat y_{s,1}}{g_1}+\frac{i_{1,m}}{\beta g_1 \sin\theta_2}-\frac{i_{in,m} \cot\theta_2}{\beta g_1 g_4 \sin\theta_1},\\
\hat y_2'=&-\frac{g_1 \cot\theta_1}{g_4}\hat x_{in}-\frac{g_1}{g_4}\hat y_{in}+\frac{i_{in,m} g_1}{\beta g_4 \sin\theta_1}\nonumber\\
&+\frac{1}{\sqrt{3\gamma}} \sqrt{\frac{i_{1,m}}{\beta g_1 \sin\theta_2}-\frac{i_{in,m} \cot\theta_2}{\beta g_1 g_4 \sin\theta_1}+\left(\frac{\cot\theta_1 \cot\theta_2}{g_1 g_4}-\frac{g_4}{g_1}\right)\hat x_{in}+\frac{\cot\theta_2}{g_1 g_4}\hat y_{in}-\frac{\hat y_{s,1}}{g_1}+\hat y_{s,2}}. 
\label{y'2}
\end{align}
\end{widetext}
Here, as in the previous section, we replaced the operators of photocurrents with c-numbers corresponding to the results of the given measurement: $i_{1,m}$ and $i_{in,m}$. In contrast to Eqs. (\ref{x'2_1})-(\ref{y'2_1}) for a Gaussian cluster, due to a non-Gaussian resource a square root in Eq. (\ref{y'2}) for $y$-quadrature arises that determines the transformation error. Note that, as for the teleportation \cite{Zinatullin2}, due to the large displacement $\alpha>0$, it is necessary to take into account only positive values of the square root. To simplify further equations, we introduce a new notation:
\begin{align}
I_m=\frac{i_{1,m}}{\beta g_1 \sin\theta_2}-\frac{i_{in,m} \cot\theta_2}{\beta g_1 g_4 \sin\theta_1}.
\end{align}

We can decompose the square root in Eq. (\ref{y'2}) in a series in terms of
\begin{align}
\frac{1}{I_m}\left[\left(\frac{\cot\theta_1 \cot\theta_2}{g_1 g_4}-\frac{g_4}{g_1}\right)\hat x_{in}+\frac{\cot\theta_2}{g_1 g_4}\hat y_{in}-\frac{\hat y_{s,1}}{g_1}+\hat y_{s,2}\right], \nonumber
\end{align}
keeping only the first term in the expansion
\begin{align}
\hat y_2'=&-\frac{g_1 \cot\theta_1}{g_4}\hat x_{in}-\frac{g_1}{g_4}\hat y_{in}+\frac{i_{in,m} g_1}{\beta g_4 \sin\theta_1}
+\sqrt{\frac{I_m}{3\gamma}}\nonumber\\
&+\frac{1}{\sqrt{12\gamma I_m}}
\bigg[\left(\frac{\cot\theta_1 \cot\theta_2}{g_1 g_4}-\frac{g_4}{g_1}\right)\hat x_{in}+\frac{\cot\theta_2}{g_1 g_4}\hat y_{in} \nonumber\\
&-\frac{\hat y_{s,1}}{g_1}+\hat y_{s,2}\bigg].
\end{align}
The termination of the series is correct under the assumption that all moments of the expansion parameter are small. For Gaussian input states, it suffices to satisfy the inequalities
\begin{align}
3 \gamma \alpha^2\gg &
\left(\frac{\cot\theta_1 \cot\theta_2}{g_1 g_4}-\frac{g_4}{g_1}\right) \langle\hat x_{in}\rangle+
\left(\frac{\cot\theta_2}{g_1 g_4}\right) \langle\hat y_{in}\rangle, \\
(3 \gamma \alpha^2)^2\gg &
\left(\frac{\cot\theta_1 \cot\theta_2}{g_1 g_4}-\frac{g_4}{g_1}\right)^2 \langle\hat x_{in}^2\rangle \nonumber\\
&+2\left(\frac{\cot\theta_1 \cot\theta_2}{g_1 g_4}-\frac{g_4}{g_1}\right) \left(\frac{\cot\theta_2}{g_1 g_4}\right) \langle\hat x_{in}\rangle \langle\hat y_{in}\rangle \nonumber\\
&+\left(\frac{\cot\theta_2}{g_1 g_4}\right)^2 \langle\hat y_{in}^2\rangle+
\frac{\langle\hat y_{s,1}^2\rangle}{g_1^2}+
\langle\hat y_{s,2}^2\rangle.
\end{align}
Note that this requirement limits the protocol’s applicability. Below, we will discuss in detail how significant this limitation is.

Thus, after measurements over the oscillators in the input and first channels, the quadratures of the second oscillator take a form
\begin{align}
\begin{pmatrix}
\hat{x}'_2\\
\hat{y}'_2
\end{pmatrix}=&
\begin{pmatrix}
1 & 0\\
\frac{1}{\sqrt{12 \gamma I_m}} & 1
\end{pmatrix} 
\Bigg[
\begin{pmatrix}
\frac{\cot\theta_1 \cot\theta_2}{g_1 g_4}-\frac{g_4}{g_1} & \frac{\cot\theta_2}{g_1 g_4}\\
-\frac{g_1 \cot\theta_1}{g_4} & -\frac{g_1}{g_4}
\end{pmatrix}
\begin{pmatrix}
\hat{x}_{in}\\
\hat{y}_{in}
\end{pmatrix} \nonumber\\
&+
\begin{pmatrix}
-\frac{\hat{y}_{s,1}}{g_1}\\
\frac{\hat{y}_{s,2}}{\sqrt{12\gamma I_m}}
\end{pmatrix}\Bigg]+
\begin{pmatrix}
I_m\\
\frac{i_{in,m}}{\beta g_2 \sin\theta_1}+\sqrt{\frac{I_m}{3\gamma}}
\end{pmatrix}.
\label{transform_cpg}
\end{align}
Let us compare the resulting expression with the Eq. (\ref{x2}) for the transformation on a pair of Gaussian resource oscillators. One can see that the nonlinearity of the cubic phase gate leads to the appearance of an additional deformation (the matrix before the square brackets on the right side of the Eq. (\ref{transform_cpg})). At the same time, the deformation coefficient depends on the measured values of the photocurrents, which is why we cannot control it. Therefore, we need to compensate for this deformation. Otherwise, it will distort the result, and we can significantly increase the error.

The operation on the second pair of nodes does not contain nonlinearity, so it is similar to the transformation (\ref{x2}) up to weight coefficients. At the output of the scheme, the c-number components of the quadratures of the field are compensated by displacement, depending on the values of the measured photocurrents. In addition, we explore notation (\ref{theta_new}), as in the previous section. Then the transformation performed on the input oscillator has the form:
\begin{align}
\begin{pmatrix}
\hat x_{out}\\
\hat y_{out}
\end{pmatrix}&=
\begin{pmatrix}
\frac{\cot\theta_3 \cot\theta_4'-1}{g_3/g_2} & \frac{\cot\theta_4'}{g_3/g_2}\\
-\frac{g_3 \cot\theta_3}{g_2} & -\frac{g_3}{g_2}
\end{pmatrix}
\begin{pmatrix}
1 & 0\\
\frac{1}{\sqrt{12 \gamma I_m}} & 1
\end{pmatrix} \nonumber\\
&\times\left[
\begin{pmatrix}
\frac{\cot\theta_1 \cot\theta_2'-1}{g_1/g_4} & \frac{\cot\theta_2'}{g_1/g_4}\\
-\frac{g_1 \cot\theta_1}{g_4} & -\frac{g_1}{g_4}
\end{pmatrix}
\begin{pmatrix}
\hat{x}_{in}\\
\hat{y}_{in}
\end{pmatrix}+
\begin{pmatrix}
-\frac{\hat{y}_{s,1}}{g_1} \\
\frac{\hat{y}_{s,2}}{\sqrt{12\gamma I_m}}
\end{pmatrix}\right] \nonumber\\
&+
\begin{pmatrix}
-\frac{\hat y_{s,3}}{g_3}\\
\hat y_{s,4}
\end{pmatrix}.
\label{out1}
\end{align}
Now let us find if it is possible to compensate for the deformation that occurs due to the presence of the cubic phase state. This deformation leads to a distortion of the oscillator basis at the output of the first part of the scheme. Therefore, knowing the measurement results in the first part of the scheme, we have the opportunity to remove this deformation by correcting the measurement basis in the second part of the scheme. For making it, we rewrite the Eq. (\ref{out1}), including the deformation in the matrix of the second part of the protocol:
\begin{align}
\begin{pmatrix}
\hat x_{out}\\
\hat y_{out}
\end{pmatrix}&=
\begin{pmatrix}
\frac{g_2}{g_3}\left[\left(\cot\theta_3+\frac{1}{\sqrt{12\gamma I_m}}\right) \cot\theta_4-1\right] & \frac{g_2}{g_3}\cot\theta_4\\
-\frac{g_3}{g_2}\left(\cot\theta_3+\frac{1}{\sqrt{12\gamma I_m}}\right) & -\frac{g_3}{g_2}
\end{pmatrix} \nonumber\\
&\times\left[
\begin{pmatrix}
\frac{\cot\theta_1 \cot\theta_2-1}{g_1/g_4} & \frac{\cot\theta_2}{g_1/g_4}\\
-\frac{g_1 \cot\theta_1}{g_4} & -\frac{g_1}{g_4}
\end{pmatrix}
\begin{pmatrix}
\hat{x}_{in}\\
\hat{y}_{in}
\end{pmatrix}+
\begin{pmatrix}
-\frac{\hat{y}_{s,1}}{g_1} \\
\frac{\hat{y}_{s,2}}{\sqrt{12\gamma I_m}}
\end{pmatrix}\right] \nonumber\\
&+
\begin{pmatrix}
-\frac{\hat y_{s,3}}{g_3}\\
\hat y_{s,4}
\end{pmatrix}.
\label{out2}
\end{align}
If we introduce the new phase
\begin{align}
\cot\theta'_3=\cot\theta_3+\frac{1}{\sqrt{12\gamma I_m}},
\end{align}
then the transformation over the input oscillator will be determined by the matrix (\ref{matrix_U}) depending on $\theta'_3$. In other words, according to the measurements results on the input and the first resource oscillators, we can adjust the phase $\theta_3$ to compensate for the additional deformation. Thus, we can perform the given operation without additional distortion. As a result, in the modified scheme, the transformation performed on the input state is the same as the transformation to Eq. (\ref{out0}), but it has a different error:
\begin{align}
\begin{pmatrix}
\hat x_{out}\\
\hat y_{out}
\end{pmatrix}=&
U(\theta_1,\theta_2',\theta_3',\theta_4')
\begin{pmatrix}
\hat{x}_{in}\\
\hat{y}_{in}
\end{pmatrix}
+
\delta \hat {\bf e}(\theta_3',\theta_4').
\end{align}
Here the matrix $U$ is defined by the Eq. (\ref{matrix_U}), and the error is given by
\begin{align}
\delta \hat {\bf e}_0(\theta_3',\theta_4')=&
\begin{pmatrix}
\frac{\cot\theta_3' \cot\theta_4'-1}{g_3/g_2} & \frac{\cot\theta_4'}{g_3/g_2}\\
-\frac{g_3 \cot\theta_3'}{g_2} & -\frac{g_3}{g_2}
\end{pmatrix}
\begin{pmatrix}
-\frac{\hat{y}_{s,1}}{g_1}\\
\frac{\hat{y}_{s,2}}{\sqrt{12\gamma I_m}}
\end{pmatrix} \nonumber\\
&+
\begin{pmatrix}
-\frac{\hat{y}_{s,3}}{g_3}\\
\hat{y}_{s,4}
\end{pmatrix}.
\end{align}
Thus, the conversion error depends not only on the phases $\theta_3'$ and $\theta_4'$, but also on the measured value $I_m$.

We have found that the considered scheme with a cubic phase gate implements the same transformation as the scheme with a Gaussian cluster considered in the previous section. Thus, we do not need to justify the arbitrariness of this transformation. However, the errors of these two operations differ significantly from each other. Next, we compare the computation errors of the two schemes and evaluate the limitations of the scheme with a cubic phase gate.

\subsection{Error of transformation with cubic phase gate}

Let us now investigate how the error of computation in the modified scheme has changed. As in the previous section, we assume the Gaussian resource states to be squeezed equally ($\langle \hat y^2_{s,j} \rangle \equiv \langle \delta \hat y^2_s \rangle$ for $j\in1,2,3,4$). Then, the variances of the error vector are following:
\begin{align}
\langle \delta \hat {\bf e}^2_0 \rangle=&
\begin{pmatrix}
\frac{1}{g_1^2}\left(\frac{\cot\theta_3' \cot\theta_4'-1}{g_3/g_2}\right)^2+\frac{1}{12 \gamma I_m}\left(\frac{\cot\theta_4'}{g_3/g_2}\right)^2+\frac{1}{g_3^2}\\
\frac{1}{g_1^2}\left(\frac{g_3 \cot\theta_3'}{g_2}\right)^2+\frac{1}{12 \gamma I_m}\frac{g_3^2}{g_2^2}+1
\end{pmatrix} \nonumber\\
&\times\langle \delta \hat y^2_s \rangle.
\label{err1}
\end{align}
Comparing Eqs. (\ref{de_0_2}) and (\ref{err1}), one can see that the second term in the expressions for quadrature errors is smaller in $12 \gamma I_m$. The average value of $I_m$ is proportional to the displacement $\alpha$ of resource state quadrature, therefore, with a sufficient displacement $\alpha$, we can significantly decrease the contribution to the error from these terms.

\begin{figure}[t]
\begin{center}
\includegraphics[width=85mm]{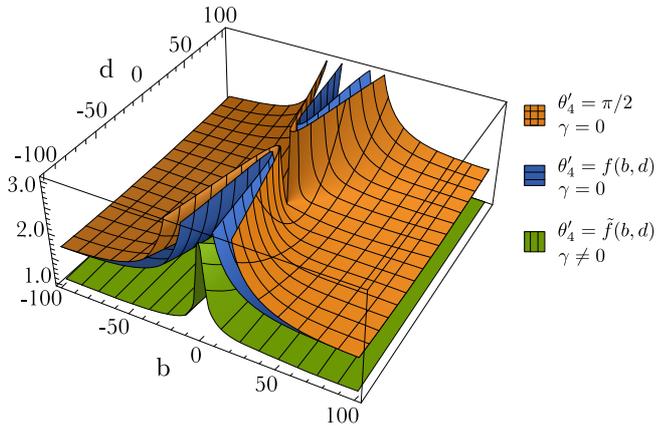}
\caption{The distribution of single-mode transformation errors $||\langle \delta \hat {\bf e}^2_0 \rangle||_{\infty}/\langle \delta \hat y^2_s \rangle$. The graph demonstrates three error distributions depending on the implemented operation, i.e. it depending on $b$ and $d$ (see Eq. (\ref{univer1})). All three distributions are calculated for optimized weight coefficients ($g_1=g_2=5$, $g_3=g_4=4$). The orange and blue surfaces correspond to errors for the scheme without a cubic phase gate ($\gamma=0$): the orange surface corresponds to the case of a fixed phase value $\theta_4'=\pi/2$, the blue surface corresponds to the case when the angle $\theta_4'$ optimization is performed ($\theta_4'=\tilde f(b,d)$). The green surface corresponds to the error of the scheme using the cubic phase gate ($\gamma\neq 0$) and $\theta_4'$ is being optimized.}
\label{cub_phase_opt}
\end{center}
\end{figure}

As in the previous section, we consider the norm  $||\cdot||_{\infty}$ as a measure of errors. We estimate the value of $I_m$ as its average value, i.e. $\langle I_m\rangle \approx 3\gamma \alpha^2$. We used a relatively small cubic phase gate coefficient $\gamma=0.1$ \cite{Yukawa, Miyata} and the displacement $\alpha=5\sqrt{5}$ (i.e. $12\gamma I_m=45$) in the calculations. This displacement satisfies condition $\alpha^2\gg\langle \hat x_{s,2}^2 \rangle$ required for the correct operation of the protocol and is implemented in practice. Fig. \ref{cub_phase_opt} demonstrates a comparison of error surfaces for a scheme without and with a cubic phase gate. One can see, that the error of the scheme with a cubic phase gate turns out to be smaller for the entire range of transformations. In addition, it suppresses the increase of the error in the vicinity of $b=0$. Thus, the inclusion in the cluster of a non-Gaussian resource gotten using a cubic phase gate further reduces the computation error.

Let us remind that the proposed protocol can operate under the condition of low nonlinearity of the cubic phase gate. We can compensate for a small value of $\gamma$ by a large value of displacement $\alpha$. This is an important advantage, since increasing the transformation coefficient $\gamma$ of the cubic phase gate is a difficult experimental problem.

\section{Evaluation of the optimization efficiency of one-way quantum computation} \label{Seff}

\begin{figure*}[t]
\begin{center}
\includegraphics[scale=0.87]{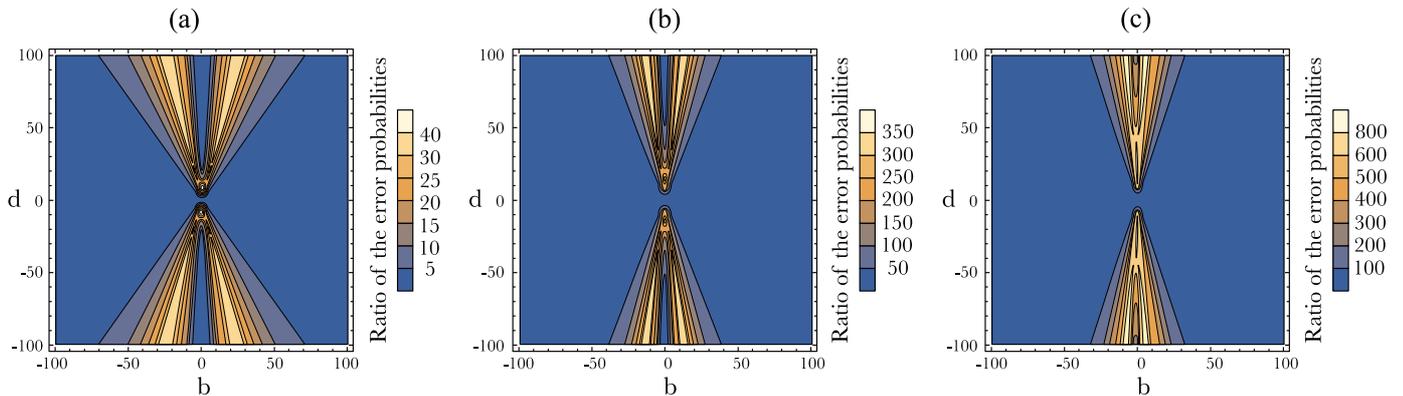}
\caption{The ratios of the error probabilities $P_{err}(x_{er},y_{er})/P_{err}(x^{opt}_{er},y^{opt}_{er})$, depending on $b$ and $d$ (see Eq. (\ref{univer1})). Here $P_{err}(x_{er},y_{er})$ is error probabilities of non-optimize computations, $P_{err}(x^{opt}_{er},y^{opt}_{er})$ is error probabilities of optimize computations. On the diagram: a) optimization is performed by weight coefficients ($g_4=g_3=4$, $g_2=g_1=5$ and $\theta_4'=\pi/2$); b) optimization is performed by weight coefficients and the phase of the homodyne detector; c) full optimization is performed, and cubic phase gate is used.}
\label{Fig_prob}
\end{center}
\end{figure*}

In the previous sections, we have shown that one-way computations on a four-node cluster state can be optimized by choosing weight coefficients as well as the use of non-Gaussian transformations. The optimization leads to decreasing of the computation errors. Now, we need to discuss how effective these optimizations are. What is the gain in decreasing errors if we apply all the proposed optimizations?

To answer these questions, we first need to understand how the resulting errors can be compared with each other and what advantages we can expect from the proposed optimization. To do this, we recall the quantum error correction procedure. In one-way quantum computations, the error displaces the quadrature of the state under computation by a small value proportional to the squeezed quadrature variance of the resource oscillator. In \cite{GKP}, the authors proposed a method for correcting errors of small quadrature displacements using the so-called GKP (Gottesman, Kitaev, Preskill) states. In \cite{Menicucci1}, this method was theoretically applied to the problem of error correction of one-way computation. In \cite{Korolev_2022}, we refined the error correction method for one-way computations taking into account the noise inherent in the error correction procedure itself, i.e. non-ideal GKP states. Omitting all the theoretical details, we can conclude the probability that the error has not been corrected 
is as follows
\cite{Korolev_2022}:
\begin{align}
P_{err}(x_{er}&,y_{er})=1-\mathrm{erf}\left(\frac{\sqrt{\pi}}{2\sqrt{2}\sqrt{\langle \delta \hat{y}^2_s\rangle \left(x_{er}+\frac{\sqrt{5}+1}{2}\right)}}\right) \nonumber\\
&\times\mathrm{erf}\left(\frac{\sqrt{\pi}}{2\sqrt{2}\sqrt{\langle \delta \hat{y}^2_s\rangle \left(y_{er}+\sqrt{5} +1\right)}}\right),
\end{align} 
where $x_{er}\langle \delta \hat{y}^2_s\rangle$ is the error variance of x-quadrature of the output target oscillator, $y_{er}\langle \delta \hat{y}^2_s\rangle$ is the error variance of the y-quadrature. The arguments of the functions $\mathrm{erf}$ are determined by two factors: the variances of the transformation errors of each quadrature (the first terms in the denominators of the arguments) and the error added when performing error correction. The latter factor is also the sum of two contributions: the error from the performing of operation $SUM(1)$ and the error from the broadening of the GKP state peaks. Note that the order of the correction procedure determines the distinction of $x$- and $y$- quadrature errors. The error function $\mathrm{erf}(1/z)$ is monotonically decreasing, so the greater the value of error variance of quadratures, the more likely the errors have not been corrected.

As can be seen from the definition of the function $P_{err} (x_{er}, y_{er})$, it reveals the quality of computation and characterizes its scale. Accordingly, it is convenient to utilize this function as a measure for comparing optimized and non-optimized computations and for evaluating the efficiency of the optimization procedure. Fig. \ref{Fig_prob} demonstrates the ratios of the error probabilities $P_{err}(x_{er},y_{er})/P_{err}(x^{opt}_{er},y^{opt}_{er})$, where $P_{err}(x_{er},y_{er})$ is the error probability of non-optimized computation, $P_{err}(x^{opt}_{er},y^{opt}_{er})$ is the error probability of optimized computation. Non-optimized computations correspond to ones on the unweighted cluster state with $\theta_4'=\pi/2$. Optimized computations correspond to ones discussed in Fig. \ref{cub_phase_opt}: a) optimization by weight coefficients ($g_4=g_3=4$, $g_2=g_1=5$) and $\theta_4'=\pi/2$; b) optimization by weight coefficients and by the phase of the homodyne measurement ($\theta_4'=f(b,d)$); c) optimization by weight coefficients, by the phase of homodyne detector, and using a cubic phase gate. All graphs are calculated for squeezing of $-15$ dB. When we perform the optimization only by weight coefficients, the error probability for some transformations becomes $45$ times less. If, in addition, we perform optimization via the phases of the homodyne detector, the gain for some operations is $400$ times. When we use a cubic phase gate and perform full optimization, the error probability for some transformations is $900$ times smaller. Note that for resource oscillators with less squeezing, the benefit from the optimization procedure is even more significant.

Thus, the proposed optimization works very effectively. We can decrease the error probability in the results of computations after the correction procedure by several orders of magnitude. This means that the optimized computation scheme is more fault-tolerant. The fault-tolerant universal quantum computation in the proposed scheme requires less squeezing than has been suggested earlier \cite{Menicucci1}.


\section{Conclusion}

In the presented work, we have shown that varying the weight coefficients of the cluster state, which used as a resource for computations, one can decrease the error of arbitrary single-mode Gaussian transformations. 
In real experiments, the squeezing resource is not infinite. Its proper distribution in the cluster is required. 
We estimated the upper value of the weight coefficients  that could be obtained with the current experimental capabilities. 
We have shown that ratios of weight coefficients play a significant role in decreasing the error. Proper distribution of weight coefficients allows us to decrease the error for most of the single-mode Gaussian operations whilst spending less energy.

For non-universal operations,
it is possible to select the cluster state configuration that provides minimal computation error. Generally, the problem of multidimensional optimization is extremely complex. Its complexity is determined both by the dimension of the cluster and by the infinite dimensions of the transformation space. Nevertheless, it is possible to select a weight coefficient which provides a minimal error for most of the operations.

Another useful tool is optimization by phases of homodyne measurements. Unlike the weight coefficients, we can choose the optimal phases for each specific operation. This strategy allows us to decrease the computation error without using any additional resources.

We have shown that the inclusion of non-Gaussian nodes prepared by cubic phase gates into the resource cluster state can further decrease the transformation error. For proper work of the protocol, we need to make relatively small displacements of the squeezed state before applying the cubic phase gate. These displacements can be easily implemented in practice. It should be noted that the practical implementation of cubic phase gate is still a challenge for experimentalists. However, the generation of the cubic phase states has recently been demonstrated in the microwave frequency range \cite{Kudra}.  There is also an active search for suitable systems for the implementation of non-Gaussian gates in optics \cite{Yukawa,Marshall,Yanagimoto,Asavanant,Miyata,Konno}.

We have demonstrated the effectiveness of our optimization methods. We have shown that it is possible to significantly decrease the probability of wrong error correction using the proposed optimization methods. This, makes our scheme more fault-tolerant. Thus, the considered methods can give a significant benefit for arbitrary single-mode Gaussian transformations.

It is important, to note that the optimization procedure proposed by us does not depend on the way the cluster state was generated and on the encoding of the input states. Regardless of the available experimental resources, it is possible to optimize the scheme to minimize the quantum computation error.

In future works, we intend to generalize our proposed optimization method to reduce errors in two-mode Gaussian and non-Gaussian transformations. Since all these transformations are needed to implement a universal quantum computer \cite{Lloyd}, their optimization will help make quantum computation more tolerant to errors. It will help reduce the requirements on squeezing for the used resources.

\vspace{0.5 cm}

Sec. \ref{Sgaus} was supported by the Theoretical Physics and Mathematics Advancement Foundation "BASIS" (Grant No. 21-1-4-39-1). Sec. \ref{Snongaus} and \ref{Seff} was supported by the Russian Science Foundation (Grant No. 22-22-00022).
 


 \end{document}